\documentstyle[12pt,frascatiphys,epsfig]{article}
\begin{document}
\title{ 
CONTROVERSIES ON AND A REASONING FOR EXISTENCE OF
THE LIGHT $\sigma$-PARTICLE
}
\author{
Shin Ishida        \\
{\em Atomic Energy Research Institute, College of Science and Technology}\\
{\em Nihon University, Kanda-Surugadai, Chiyoda-ku, Tokyo 101-0062, Japan} 
}
\maketitle
\baselineskip=14.5pt
\begin{abstract}
The light $\sigma$-particle is, regardless of the strong criticism, reviving
recently due to the works done from various sides. I review essential
points of the controversies (especially related to our works) and of their 
answer:
Conventionally a large concentration of the iso-scalar $S$-wave 
2$\pi$ events below 1~GeV (being, correctly, due to the 
$\sigma$-production), which is observed in most of production 
processes, is interpreted as a mere background from the viewpoint of,
so called, universality argument.
However, I show, by resorting to a simple field theoretical model,
that the argument is not correct and the production process has
``its own value'' independent of the scattering process.
Thus it is suggested that the present index 
``$f_0$(400-1200) or $\sigma$'' in PDG'98
is to be changed as  ``$\sigma$(400-800)'' 
in the PDG 2000.
\end{abstract}
\baselineskip=17pt
\section{Introduction}
\subsection{Recent short history of the $\sigma$-particle
            and the related works}
Recently the many works,\footnote{
In this talk I refer only to the recent works after 1980. As for the 
old references on the $\sigma$-particle see the works referred in 
S.Ishida et al.'96 in References.\cite{ref1}
} suggesting existence of 
the light $\sigma$-particle both theoretically and phenomenologically,
had been published and the $\sigma$-particle was revived in the newest
lists of PDG'96 and '98 after missing more than two decades, although
still with an obscure index ``$f_0(400$-$1200)$ or $\sigma$''. Among
them our group\footnote{
The members are S.Ishida, T.Komada and H.Takahashi(Nihon University);
M.Y.Ishida(Tokyo Institute of Technology); K.Takamatsu(Miyazaki University);
and T.Ishida and T.Tsuru(KEK).
} of collaboration had also given rather
strong evidences for its existence through a series of papers,\cite{ref1} while
received a serious criticism. In this talk I shall summarize the essential
points of controversies and explain our answers to the criticism clearly,
leading to the suggestion as is given in the abstract:

From early 1980's the importance of the $\sigma$ in relation to the 
dynamical chiral symmetry breaking had been stressed by the works.\cite{ref2}
Possible evidences suggesting for existence of the $\sigma$ in production
processes had been given in the works.\cite{ref1} The reanalyses of the 
$\pi\pi$ scattering phase shifts, leading to the rather strong evidences
for the $\sigma$ were done by the works.\cite{ref4}

The results of our series of works were reported in the several 
occasions,\cite{ref1} of which criticisms are found in the 
references.\cite{ref6,ref7} 
Some useful arguments and discussions, which make the 
crucial points clear, were given in the references.\cite{ref8}
\subsection{Outline of controversies}
First possible evidence for direct $\sigma$-production, which was obtained
in a proton proton central collision process at 450GeV/c
\begin{equation}
pp\rightarrow p_fX^0p_s,\ \ X^0\rightarrow n\gamma,\ \ X^0=(\pi\pi )
\end{equation}
was reported\cite{ref1} by the GAMS KEK-subgroup at Manchester, Hadron'95.
The obtained $(\pi^0\pi^0)$ mass spectra are given in Fig. 1. They fitted
the spectra by the Variant Mass and Width(VMW) method, 
representing the invariant production amplitude as a coherent sum of 
Breit-Wigner amplitudes of resonances, $X^0=f_0(975),f_2(1275)$ and $f_c(500)$.
\begin{figure}[t]
 \epsfysize=4.0 cm
 \centerline{\epsffile{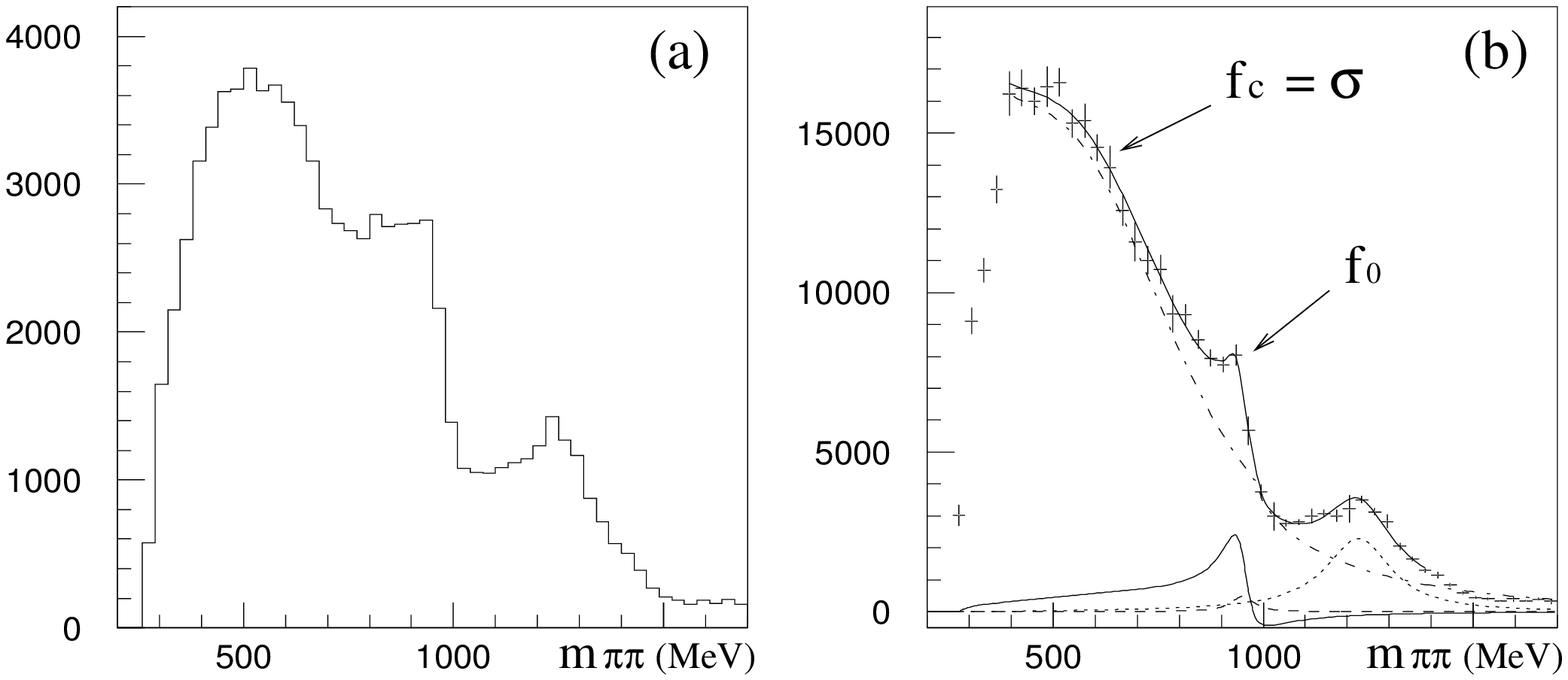}}
 \caption{\it
      $\pi^0\pi^0$ mass spectra in GAMS $pp$ central collision 
      experiment, reported in Hadron'95. (a) not corrected for
      acceptance (b) corrected
    \label{had95} }
\end{figure}
There the huge concentration of $S$-wave events below 1 GeV, to which 
similar spectra had also been found in other experiments\cite{ref9} 
and taken as the
mere background before, was interpreted as being due to 
$f_c(500)=\sigma (500)$-particle production.

However, this interpretation of $\sigma$ production was severely 
criticized\cite{ref6} in the summary talk of the Hadron'95 from the 
so-called ``Universality Argument.'' It says ``claims of a narrow 
$\sigma (500)$ in the GAMS results cannot be correct as''\\
$\bigcirc\hspace{-0.3cm}1$\hspace{0.3cm}  No $\sigma$ is seen in 
                                    the $\pi\pi$ scattering.\\
$\bigcirc\hspace{-0.3cm}2$\hspace{0.3cm}  Unitarity demands 
            the production amplitude ${\cal F}$ to be consistent 
            with the scattering amplitude ${\cal T}$.\\
Due to this serious criticism the GAMS group himself had taken a very cautious
attitude\cite{ref10} on the $\sigma$-particle to state formally that
``In summary the analysis of $\pi^0\pi^0$ system $\cdots$ confirms a large
concentration of $S$-wave events below 1 GeV, which interferes with 
$f_0(980)$ destructively $\cdots$. This would be compatible with a broad 
$S$-wave state $\cdots$ but its coference with the known $\pi\pi$-scattering
phase shifts is still the object of controversy that bears basic 
non-perturbative QCD concepts.''

On the other hand our group had reanalyzed,\cite{ref1} in replying to the
criticism $\bigcirc\hspace{-0.3cm}1$, the $\pi\pi$ phase shifts by using
the Interfering (Breit-Wigner) Amplitude (IA) method which satisfies the 
elastic unitarity automatically and shown that the $\sigma$-particle actually 
exists. Furthermore, our group had also investigated,\cite{ref12} in replying
to the criticism $\bigcirc\hspace{-0.3cm}2$, the relation between the
scattering and the production amplitudes and shown our ${\cal F}$ in the 
VMW method and ${\cal T}$ in the IA method satisfy consistently the unitarity
of $S$-matrix. Meanwhile, there have been opened some useful 
arguments\cite{ref8}
to make clear the critical points. Through the above processes
I believe that now the answers to all the criticisms have been given.
\section{$\pi\pi$-scattering amplitude and reanalyses of phase shifts}
We made recently a reanalysis\cite{ref1} of the old CERM-Munich '73 and '74
data of $\pi\pi$ phase shifts and found a strong evidence for existence of 
$\sigma$-particle. There we applied the IA-method, which satisfies the elastic
unitarity automatically and is parametrized only in terms of physically
meaningful quantities as masses and widths of resonances. In a simple case of
one $(\pi\pi )$-channel and two resonant $(\sigma$ and $f_0)$ particles the
partial $S$-wave $S$-matrix in the IA method is given as follows:
\begin{eqnarray}
S &=& S^{\rm Res}S^{\rm BG},\ \ S^{\rm Res}=S^\sigma S^{f_0},\nonumber\\
S^{(i)} &=& e^{2i\delta^{(i)}},\ \ 
         \delta =\delta^\sigma +\delta^{f_0}+\delta^{\rm BG},
\end{eqnarray}
where $S^{\rm Res}(S^{\rm BG})$ corresponds to $S$-matrix in the case of
pure resonant (background) scattering and the $\delta^{(i)}$ represent
the phase shifts due to the respective pure  scattering cases. The unitarity
of total $S$-matrix $S$ is reduced to the unitarity of each 
``component $S$-matrix'' $S^{(i)}$;
\begin{eqnarray}
SS^\dagger &=& S^\dagger S=1 \leftarrow  
   S^{(i)}S^{(i)\dagger} = S^{(i)\dagger} S^{(i)}=1.
\end{eqnarray}
The scattering amplitude $a(S\equiv 1+2ia;\ a=\rho {\cal T}(s))$ 
due to resonances $a^{\rm Res}$
is given as 
\begin{eqnarray}
a^{\rm Res} &=& a_{\rm BW}^\sigma +a_{\rm BW}^f
               +2ia_{\rm BW}^\sigma a_{\rm BW}^f,
\label{eq4}
\end{eqnarray}
where $a_{\rm BW}^{\sigma (f)}$ represents the Breit-Wigner amplitude of the 
$\sigma (f)$ resonance ($a_{\rm BW}^\sigma\equiv \rho g_\sigma^2
/(m_\sigma^2-s-i\rho g_\sigma^2)$ etc., $\rho =\sqrt{1-4m_\pi^2/s}/16\pi)$). 
The last term in the r.h.s. of Eq.(\ref{eq4})
represents an ``interference'' between the $\sigma$ and $f$ (B.W.) amplitudes.
The physical reason for obtaining the different result even with using the 
same experimental data is our introduction of ``negative background phase''
$\delta_{\rm BG}$ of hard core type
\begin{eqnarray}
\delta_{\rm BG} &=& -|{\bf p}_1|r_c,
\end{eqnarray}
where $|{\bf p}_1|=\sqrt{s/4-m_\pi^2}$ is the pion momentum in the $2\pi$
CM system and the $r_c$ a parameter. The physical origin of the  
$\delta_{\rm BG}$ is able to be reduced\cite{ref13} to the compensating
repulsive interaction guaranteed by the chiral symmetry,\cite{ref14} 
and it is describable
quantitatively in the framework of linear $\sigma$ model including the 
$\rho$-meson contribution.\cite{ref15}
\begin{table}[b]
\caption{
Comparison between the fit with $r_c\neq 0$
and with $r_c= 0$ in our PSA. The latter corresponds
to the conventional analyses thus far made.
}
\begin{center}
\begin{tabular}{|c|c||c|}
\hline
\hline
 & $r_c\neq 0(\chi^2/N_f=23.6/30)$
               & $r_c= 0(\chi^2/N_f=163.4/31)$    \\
\hline
    & $\delta^{\rm tot}=\delta_{f_0(980)}
      +[\delta_{\sigma (600)}+\delta_{\rm BG}
        ]^{\rm pos.}$
    & $\delta^{\rm tot}=\delta_{f_0(980)}
      +[\delta_{\rm BG}^{\rm pos.}]$ \\
    & $\sigma (600)$ 
     &  ``$\sigma$" (equivalent to $\epsilon (900)$)\\
\hline 
$m_\sigma$ & $585\pm 20({\tiny 535\sim 675})$  &  920\\ 
$\Gamma_\sigma^{(p)}$ & $385\pm 70$ & 660\\
$\sqrt{s_{\rm pole}}/$MeV & $(602\pm26)-i(196\pm 27)$ 
                    & 970-i320  \\
$r_c$   &  3.03$\pm$0.35 GeV$^{-1}$  &  -- \\
   ~    &  (0.60$\pm$0.07 fm)        & (--)\\
\hline
\end{tabular}
\end{center}
\label{tab:conventional}
\end{table}
The results of our re-analyses are given in Fig.~2(a), while
in Table~\ref{tab:conventional} I compare
the essential points and the results of our analysis with those of 
the conventional one.\cite{ref16} 
In our analysis the introduction
of repulsive $\delta_{\rm BG}$ with $r_c\sim 3$GeV${}^{-1}$
(0.60fm, about the structural size of pion)
plays a crucial role for the existence of $\sigma$(600).
The sum of the large attractive $\delta_{\sigma (600)}$, 
contribution due to $\sigma$(600),
and the large repulsive $\delta_{\rm BG}$ gives a small
positive phase shift,
which was treated,
in the conventional analysis,
as a background (or broad $\epsilon ($900))
contribution $[\delta_{\rm BG}^{\rm pos.}]$.
Note that the fit with $r_c$=0 in our analyses corresponds to
the conventional analyses without the repulsive
$\delta_{\rm BG}$ thus far made.
In this case the mass and width of ``$\sigma$"
becomes large, and the ``$\sigma$"-Breit-Wigner
formula can be regarded as an  
effective range formula 
describing a positive background phase.
The corresponding pole position
is close to that of $\epsilon$(900) in Ref.~\cite{ref16}.
In this case the value of $\chi^2$ is 
$\chi^2$ = 163.4, worse by 140 than that in our best fit.

\begin{figure}[t]
 \epsfysize=8.0 cm
 \centerline{\epsffile{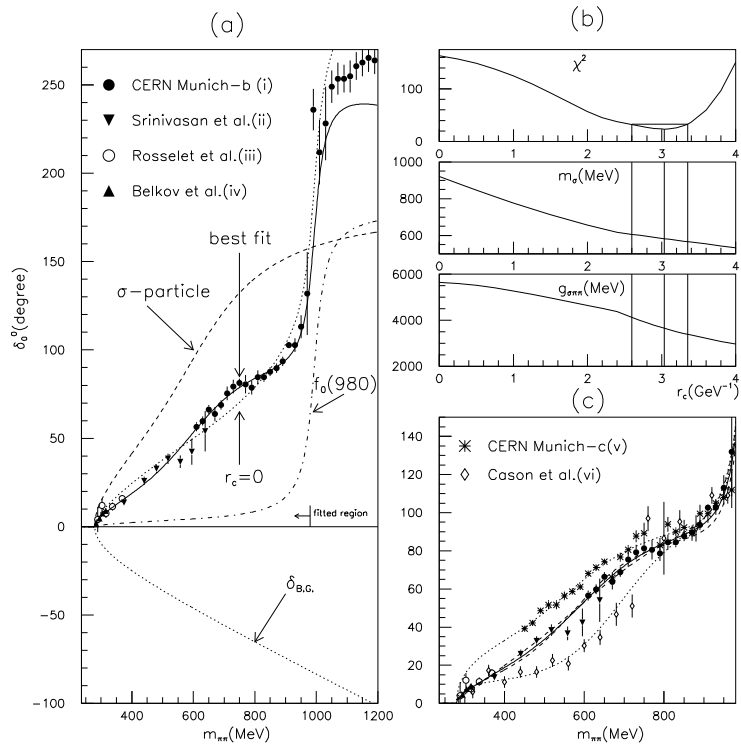}}
 \caption{\it
  $I$=0 $\pi\pi$ scattering phase shift.
  (a) Best fit to the standard $\delta_0^0$.
  The dotted line labeled
  ``$r_c$=0'' represents the conventional fit without 
  the repulsive background.
  (b) $\chi^2, M_\sigma$ and $g_\sigma$ versus $r_c$. 
  (c) Fits to the upper and lower $\delta_0^0$.
    \label{further} }
\end{figure}

\section{Production amplitude and its relation to scattering amplitude}
\subsection{General problem}
We found also some evidences\cite{ref1,ref15}
for existence of the $\sigma$-particle as an intermediate state of the
$\pi\pi$ system in the production processes\footnote{
Recently we have made a preliminary analysis of the $m_{\pi^0\pi^0}$ 
spectra in the process $p\bar p\rightarrow 3\pi^0$ observed in the crystal
barrel experiment, and found that they are reasonably well understood as 
due to production of the $\sigma$ with $m_\sigma\approx 700$ MeV and 
$\Gamma_\sigma \approx 600$ MeV in addition to the resonances considered 
there.}  
by analyzing the data obtained through the 
{\em pp} central collision experiment by
GAMS\cite{ref1,ref10} and 
the data in the 
$J/\psi\rightarrow\omega\pi\pi$ decay 
reported by DM2 collabration\cite{ref17}.
In the analyses we applied the 
Variant Mass and Width(VMW)-method,\footnote{
It was named\cite{ref18} historically after the following reason.
The mass and width of ``a" resonant particle,
which is misinterpreted as one resonance
instead of actual two overlapping resonances, 
are observed variantly depending upon the respective processes.
} 
where the production amplitude is represented by a 
sum of the $\sigma$ and $f_0$ Breit-Wigner amplitudes 
with relative phase factors
\begin{eqnarray}
\frac{r_\sigma e^{i\theta_\sigma}}{m_\sigma^2-s-i\sqrt{s}\Gamma_\sigma (s)}
+\frac{r_f e^{i\theta_f}}{m_f^2-s-i\sqrt{s}\Gamma_f (s)},
\label{eq6}
\end{eqnarray}
The general problem to be examined is whether our applied methods of analyses
are consistent with the unitarity of $S$ matrix:
The scattering amplitude ${\cal T}$ must satisfy
the elastic unitarity
and the production amplitude ${\cal F}$ must have, 
in case that the initial state has no strong phase,
 the same 
phase as ${\cal T}$:
${\cal T}\propto e^{i\delta}\rightarrow
{\cal F}\propto e^{i\delta}$
(FSI; Final-State-Interaction
theorem).
Conventionally, the more restrictive relation
between  ${\cal F}$ and  ${\cal T}$
is required on the basis of the 
``universality,"\cite{ref6,ref16}
\begin{eqnarray}
{\cal F} &=& \alpha (s){\cal T}
\label{eq:FaT}
\end{eqnarray}
with a {\it slowly varying} real function $\alpha (s)$ of $s$.
I have already shown that our ${\cal T}$ in the IA method satisfies 
the elastic unitarity automatically. The remaining problem is whether 
our ${\cal F}$ in the VMW method is consistent with the FSI theorem 
or not.

\subsection{Basic consideration}

Here I shall describe our general line of thought 
on the strong interaction of hadrons, our relevant problem. It is a 
residual interaction of QCD among color-singlet bare-hadrons, which
are the $\underline{\rm stable}$ bound states of quark and anti-quark
systems. First let us consider an old example of the strong interaction
among pions and nucleons. $\underline{\rm Before}$ knowing the quark 
physics, the $\rho$ and the $\Delta$ were $\underline{\rm resonances}$
of $2\pi$ system and $\pi N$ system, respectively, produced through the 
strong interaction among the basic pion and nucleon fields. However, presently
$\underline{\rm after}$ knowing the quark physics, the $\rho$ and the $\Delta$
should also be treated as basic fields equally as the $\pi$ and the $N$:
The stable bare particle $\bar\rho\;(\bar\Delta )$ as the bound state of 
$q\bar q\;(qqq)$ system becomes the unstable physical particle 
$\rho_{\rm phys}\;(\Delta_{\rm phys})$ after switching on the strong 
interaction among bare-particles $\bar\pi ,\bar\rho ,\bar N$ and $\bar\Delta$.
In this example an $S$-matrix $S$ consistent with the unitarity is obtained,
in the framework of (local) field theory, following the conventional procedure,
if we know\footnote{ We suppose that a theory of strong interaction among 
local hadron fields is valid as a low energy effective theory of QCD.
}
a properly supposed strong interaction Hamiltonian $H$ among basic bare fields,
which is hermitian $H^\dagger =H$:
\begin{eqnarray}
SS^\dagger =S^\dagger S &=& 1\ \ \ \leftarrow \ \ \  H=H^\dagger .
\end{eqnarray}
In our relevant problem of scalar mesons, 
we should take as basic fields the bare fields $\bar\sigma$ and $\bar f$ as well as 
the $\bar\pi$. Here we take a view-point that the $\sigma$ and the $f$
are some intrinsic quark-dynamics states(possibly to be relativistic
$S$-wave $q\bar q$ states). In this case we set up (as a simple example)
the strong interaction Hamiltonian
\begin{equation}
H_{\rm int}^{\rm scatt}=\sum_{\alpha =\sigma ,f}\bar g_\alpha\bar\alpha\pi\pi
+\bar g_{\pi\pi}(\pi\pi )^2,\ \ 
H_{\rm int}^{\rm prod}=\sum_{\alpha =\sigma ,f}\bar \xi_\alpha\bar\alpha ``P''
+\bar\xi_{\pi\pi}\pi\pi ``P'',
\label{eq7}
\end{equation}
where $\bar g$ and $\bar\xi$ are real coupling constants, ``P'' denoting a 
relevant production channel. Due to the (former) interaction (\ref{eq7})
the stable bare states $\bar\pi ,\bar\sigma$ and $\bar f$ change into the
physical states denoted as $\pi =(\bar\pi )$, and $\sigma$ and $f$ with 
finite widths. Then we can derive the scattering and production amplitudes 
following the standard procedure of quantum field theory.

\begin{figure}[t]
 \epsfysize=4.5 cm
 \centerline{\epsffile{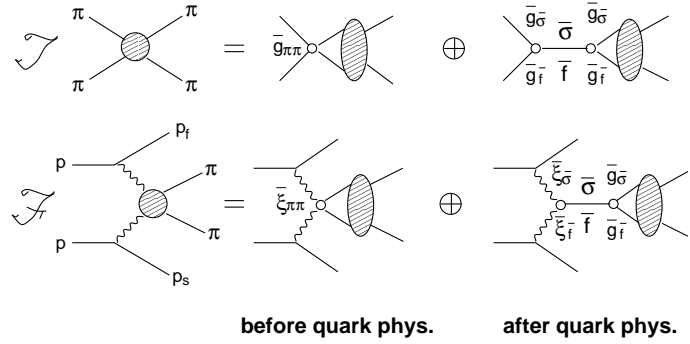}}
 \caption{\it  The mechanism for scattering amplitude ${\cal T}$ and 
      production amplitude ${\cal F}$. The latter diagram should be 
      correctly taken into account as well as the former, whereas
      only the former had been considered in the conventional treatment. 
    \label{figcomp} }
\end{figure}

The general structure of ${\cal T}$ and ${\cal F}$ is shown schematically in
Fig. \ref{figcomp}, where shaded ellipses represent the final state 
interaction of the $2\pi$ system. It is worthwhile to note that correctly 
both the mechanisms in Fig. \ref{figcomp} should be taken into account. 
As a result the ${\cal T}$ and the ${\cal F}$ are, in principle, mutually
independent quantities, reflecting the coupling constants 
$\bar g_\alpha$ and $\bar\xi_\alpha$ being so.

In the conventional treatment, where 
only the former mechanism is taken into account,
the function $\alpha (s)$ in Eq.(\ref{eq:FaT})
becomes
\begin{eqnarray}
\alpha (s) &=& \bar\xi_{\pi\pi}/\bar g_{\pi\pi}={\rm const}.
\end{eqnarray}
This leads to essentially the same $2\pi$-mass spectra in any production 
process as in the scattering process, which is evidently inconsistent with
experimental facts. 
Accordingly in the conventional analysis the $\alpha (s)$ is
assumed to have the form (which is generally {\it not varying slowly})
\begin{eqnarray}
\alpha (s) &=& \sum_n\alpha_ns^n/(s-s_0^{\cal T}),
\end{eqnarray}
introducing\footnote{
In our model the parameters $\alpha_n$ and $s_0^{\cal T}$ are 
determined from physical quantities $\bar g,\bar\xi$ and $\bar m$.
} the physically meaningless parameters $\alpha_n$, and fixing 
the value of $s_0^{\cal T}$, the zero-point of ${\cal T}$, from the 
scattering experiments. 
This procedure implies that production experiments generally
lose their values in seeking for resonant particles.
In the correct treatment considering both the former and the latter
mechanisms, the direct peak of the $\pi\pi$ mass spectra due to
the $\alpha$-particle production is to be observed in the production process,
if its production coupling constant $\bar\xi_\alpha$ is dominant, 
in conformity with our intuition.
Thus the production experiments have generally their own values independent
from the scattering experiments. 

\subsection{Justification of IA method and VMW method}
\begin{figure}[t]
 \epsfysize=2.5 cm
 \centerline{\epsffile{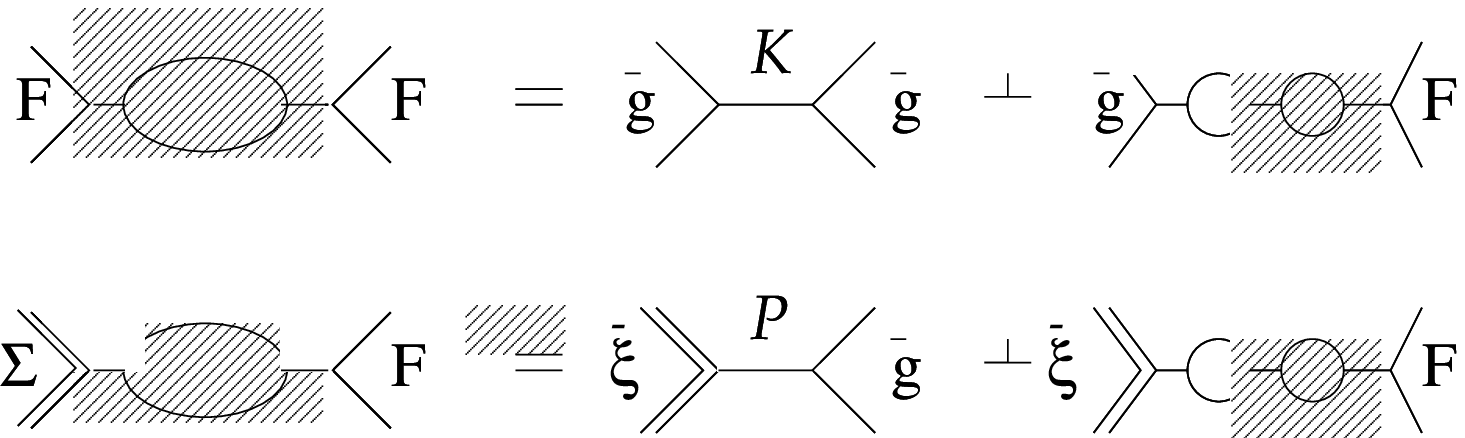}}
 \caption{
Scattering and production mechanism 
in a simple field-theoretical model of resonance dominative case.
The production amplitude is obtained, following the mechanism 
shown in the figure, by replacing the 
first $\pi\pi$-coupling constant $\bar g$ in ${\cal T}$
with the production coupling $\bar\xi$.  
The  ${\cal F}$ amplitude obtained in this 
way automatically satisfies the FSI theorem. 
 }
\label{fig:prod}
\end{figure}
In the previous work\cite{ref12} resorting to the above model 
we have derived our methods of
analyses, the IA method for ${\cal T}$ and the VMW method for ${\cal F}$, 
and shown their consistency with the FSI theorem.
The obtained formulas of the amplitudes
(derived as solutions of Schwinger-Dyson equations shown in Fig. 4)
 were\cite{ref20} 
\begin{eqnarray}
{\cal T} & = & {\cal K}/(1-i\rho {\cal K}),\ \  {\cal K}=\bar g_\sigma^2 
/(\bar m_\sigma^2 -s) + \bar g_f^2/(\bar m_f^2-s),\nonumber\\
{\cal F} & = & {\cal P}/(1-i\rho {\cal K}),\ \  {\cal P}={\cal K}
(\bar g_\sigma^2\rightarrow\bar\xi_\sigma\bar g_\sigma \ \ {\rm etc.})
\label{eq12}
\end{eqnarray}
in the ``bare-state representation.''
These formulas of ${\cal T}$ and ${\cal F}$ are rewritten\footnote{
The $r_\alpha$ and $\theta_\alpha$ in Eq.(\ref{eq6}) are 
expressed in terms of $\bar g_{\bar\alpha},\ \bar \xi_{\bar\alpha},$ 
and $\lambda_\alpha (=M_\alpha -i\rho g_\alpha^2)$ and shown to be almost 
$s$-independent except for the threshold region.
However, we must note on the following:\ \ 
In the VMW-method essentially the 
three new parameters,
$r_\sigma ,\ r_f$ and the relative phase 
$\theta (\equiv\theta_\sigma -\theta_f)$,
independent of the 
scattering process,
characterize the relevant production processes.
Presently they
are represented by the 
two production coupling constants, $\bar\xi_{\bar\sigma}$
and $\bar\xi_{\bar f}$.
Thus, among the three parameters there exists
one constraint due to the FSI-theorem.
}
into the forms of Eq.(\ref{eq4}) and Eq.(\ref{eq6}), respectively,
in the ``physical state representation.''\footnote{
Here we treat a simple case of the resonance-dominative case, including
only the virtual two-$\pi$ meson effects. In Eq.({\ref{eq12}}) we also made 
simplification by identifying the ``${\cal K}$-matrix states'' with the 
bare states. As for details see Ref. \cite{ref12}
}
 The consistency of the 
amplitudes ${\cal F}$ and ${\cal T}$ are easily seen from 
Eq.(\ref{eq12}) since ${\cal K}$ and ${\cal P}$ are real and 
their phases come only from their common denominator 
$(1-i\rho {\cal K})$. 
\section{Summary and concluding remarks}
 I have explained that our methods of analyses, the Interfering
Amplitude method for treating the $\pi\pi$ scattering process
and the VMW method for the $\pi\pi$ prodcution process (which
were effective in leading to evidences for the $\sigma$-existence)
are consistent with the unitarity of $S$-matrix.
Thus the conventional treatments along the line of universality
argument are proved to be not correct.
Accordingly I have stressed that production experiments of resonant 
particles have generally their own value independent of scattering
experiments.\\
 It is considered that confirmation of the $\sigma$-particle with
low mass and vacuum quantum number, which possibly appears in various
processes, and its $\underline{\rm right}$ treatment is crucially
important for hadron physics.\\
Finally,  on the basis of the present talk, I propose that the present 
index ``$f_0(400$-1200) or $\sigma$'' in PDG'98 is to be corrected as
$\sigma (400$-800) in the PDG future edition.

{\em 
The\ present\ speaker\ acknowledges\ deeply\ to\ the\ commitee\ of\ 
this\ workshop\ 
for\ giving\ him\ this\ nice\ opportunity\ of\ presentation.\ 
I\ should\ like\ also\ to\ express\ our\ sincere\ gratitude,\ representing\
 all\ members\ of\ our\ collaboration,\ to\ 
professor\ Montanet\ for\ his\ fair\ and\ warm\ interest\ in\ our\ works.
}
\end{document}